\documentclass[a4paper,11pt]{article}
\usepackage{pos}
\usepackage{physics} 
\usepackage{CJKutf8}
\usepackage{subcaption}
\usepackage{graphicx}
\usepackage[export]{adjustbox}
\graphicspath{{./images/}}

\newcommand{\nn}{\nonumber\\}
\newsavebox{\tempfig}

\newcommand\myscale{0.7}

\title{Measurement of the TMD soft function on the lattice using the auxiliary field representation of the Wilson line}
\ShortTitle{Measurement of the TMD soft function}

\author[a]{Anthony Francis}
\author[b]{Issaku Kanamori}
\author[a,c,d]{C.-J. David Lin}
\author*[a]{Wayne Morris}
\author[e]{Yong Zhao}

\affiliation[a]{Institute of Physics, 
  National Yang Ming Chiao Tung University,\\
  1001 Ta-Hsueh Road, Hsinchu 30010, Taiwan}

\affiliation[b]{RIKEN Center for Computational Science,\\
  7-1-26 Minatojima-minami-machi, Chuo-ku, Kobe, Hyogo 650-0047, Japan}

\affiliation[c]{Centre for High Energy Physics, Chung-Yuan Christian University,\\
  Chung-Li 32023, Taiwan}

\affiliation[d]{Centre for Theoretical and Computational Physics, National Yang Ming Chiao Tung University,\\
  1001 Ta-Hsueh Road, Hsinchu 30010, Taiwan}

\affiliation[e]{Physics Division, Argonne National Laboratory,\\
  9700 S. Cass Avenue, Lemont, IL 60439, United States}

\emailAdd{waynemorris@nycu.edu.tw}

\abstract{
  The transverse momentum dependent (TMD) soft function and Collins-Soper (CS) kernel may be obtained by formulating the Wilson line in terms of auxiliary one-dimensional fermion fields on the lattice. 
  Our computation takes place in the region of the lattice that corresponds to the “spacelike” region in Minkowski space in order to obtain the Collins soft function. 
  The matching of our result to the Collins soft function is achieved through the mapping of the auxiliary field directional vector that to the Wilson line rapidity. 
  In Euclidean space, this directional vector is complex, having a purely imaginary time component.
  We present some exploratory numerical results of our lattice calculation, and discuss the methodology employed.
}

\FullConference{The 41st International Symposium on Lattice Field Theory (Lattice 2024)\\
July 28th - August 3rd, 2024\\
University of Liverpool\\}

\begin{document}
\maketitle


\section{Introduction}

Constructing the soft function and the CS kernel within the framework of lattice QCD represents a fundamental step in exploring TMD physics through lattice-based methods. The presence of rapidity divergences in TMD calculations, including the soft function, introduces specific challenges in lattice computations due to the necessity of adapting rapidity to Euclidean space via a Wick rotation. For the soft function, we propose performing lattice calculations using Wilson lines characterized by directional vectors with purely imaginary time components, specifically $\tilde n=(in^0,\vec 0_\perp,n^3)$ \cite{Francis:2023}, which is motivated by the moving heavy-quark effective theory (mHQET) formulation in Minkowski space~\cite{jiliuliu2020}. Particularly, the auxiliary field method allows one to obtain Collins' definition of the soft function \cite{collins2011a}. Furthermore, a successful computation of the soft function will allow us to compute the CS kernel and intrinsic soft function.

This approach to formulating the Wilson line enables an analytic continuation of the Euclidean soft function to its Minkowski-space counterpart within the space-like region, as the soft function remains independent in time. The lattice implementation of the Wilson line relies on the auxiliary field representation, which can be solved iteratively in Euclidean time. However, the lattice realization of an auxiliary field propagator introduces additional subtleties that will be addressed later in this work.

We establish a clear connection between this complex Wilson line direction on a Euclidean lattice and the Minkowski-space rapidity as defined within the Collins regularization scheme \cite{collins2011a}. In essence, our lattice calculations correspond to the Minkowski-space soft function derived from Wilson lines with space-like orientations.

When the Wilson line is oriented in a time-like direction, its auxiliary field representation aligns with mHQET. Here, the Wilson line's directional vector represents the velocity of the heavy quark. 
The proposal in \cite{jiliuliu2020} suggested calculating the soft function using mHQET, treating it as the form factor of a heavy quark-antiquark pair. In this setup, the heavy quark's velocity corresponds to time-like Wilson line directions in Minkowski space. Our work is inspired by this idea but diverges by employing Euclidean directional vectors that translate to space-like Minkowski directions. Additionally, we find that attempting to compute the soft function in Euclidean space with directional vectors aligned to time-like Minkowski directions results in a divergent integral.

  \begin{figure}[b]
      \centering
      \includegraphics[width=.5\linewidth]{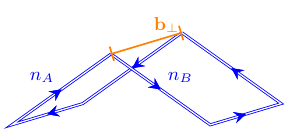}
      \caption{Sketch of butterfly shape Wilson loop.}
      \label{fig:schem}
  \end{figure}

\section{Theoretical considerations}

We conduct a perturbative investigation for both infinite and finite-length Wilson lines. For the infinite case, we establish an equivalence between computations performed in Euclidean and Minkowski spaces at one loop. Additionally, we demonstrate that the results for finite-length Wilson lines converge toward those of infinite-length lines when a ratio is employed to cancel finite-length effects.

\subsection{Infinite Wilson lines}
  \begin{figure}
    \centering
    \begin{subfigure}{0.24\textwidth}
      \includegraphics[width=\textwidth]{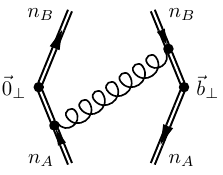}
      \caption{}
      \label{fig:Sa}
    \end{subfigure}
    \begin{subfigure}{0.24\textwidth}
      \includegraphics[width=\textwidth]{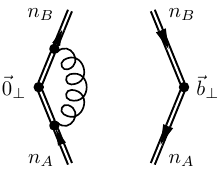}
      \caption{}
      \label{fig:Sb}
    \end{subfigure}
    \begin{subfigure}{0.24\textwidth}
      \includegraphics[width=\textwidth]{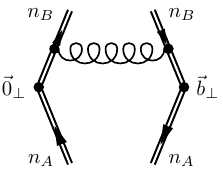}
      \caption{}
      \label{fig:Sc}
    \end{subfigure}
    \begin{subfigure}{0.24\textwidth}
      \includegraphics[width=\textwidth]{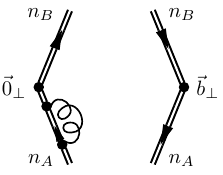}
      \caption{}
      \label{fig:Sd}
    \end{subfigure}
    \caption{Diagrams contributing at one-loop to the soft function, up to mirror diagrams.}
    \label{fig:S}
  \end{figure}

We begin by performing a one-loop computation of the soft function in Euclidean space with complex directions to establish a connection with the Minkowski space result: 
\begin{align} \tilde n_A = (in_A^0, \vec 0_\perp, n_A^3), \qquad \tilde n_B = (in_B^0, \vec 0_\perp, -n_B^3) \, . \end{align} 
For convenience, we define the ratios: $r_a = n_A^3/n_A^0$, and $r_b = n_B^3/n_B^0$. 
The computation is performed in $d$-dimensional space-time to regulate UV divergences. 
Before presenting the full result, we gain useful insights by first examining diagram (a) in Figure \ref{fig:S}: \begin{align} S_{\ref{fig:Sa}}(b_\perp, \epsilon, r_a, r_b) &= i g^2 C_F (\tilde n_{A,\mu} \tilde n_{B,\nu}) \int_{-\infty}^0 \dd s \int_{-\infty}^0 \dd t A^\mu (b_\perp+s\tilde n_A) A^\nu ( t\tilde n_B) \nn &= i g^2 C_F (\tilde n_A \cdot \tilde n_B) \int_{-\infty}^0 \dd s \int_{-\infty}^0 \dd t \int \frac{\dd^d k}{(2\pi)^d} e^{-ik(b_\perp+s\tilde n_A-t\tilde n_B)} \frac{-i}{k^2} , , \label{eq:s1aa} \end{align} where, in line 2 of Eq. (\ref{eq:s1aa}), the gluon propagator is expressed in Euclidean space. Also, $b_\perp$ refers to the transverse separation between Wilson lines.

At this point, one typically proceeds with the momentum-space computation by integrating over $s$ and $t$ to obtain the Wilson line propagator. However, these integrals are ill-defined due to the presence of a finite real part in the exponential. Focusing on the $s$-integration:

  \begin{align}
    \int_{-\infty}^0 \dd s \, e^{sn_A^0 k_4 - isn_A^3k_3}
    &=
    \left. 
      \frac{e^{sn_A^0 k_4 - isn_A^3 k_3}}
        {n_A^0 k_4 - in_A^3 k_3} 
    \right|^0_{s=-\infty}
    \overset{k_4<0}{\longrightarrow} \quad \infty \, .
  \end{align}
  Without a clear method to handle this integration, we instead proceed by performing the momentum integral in Eq. (\ref{eq:s1aa}) first. To achieve this, we use the Schwinger parametrization and complete the square in $k$:
  \begin{align}
    S_{\ref{fig:Sa}}\left(b_\perp, \epsilon, r_a, r_b\right)
    &= g^2 C_F  (\tilde n_A \cdot \tilde n_B) \int_{-\infty}^0 \dd s \int_{-\infty}^0 \dd t \int_0^\infty \dd u
    \int \frac{\dd^d k}{(2\pi)^d} 
    e^{-uk^2} 
    e^{-(b_\perp+s\tilde n_A-t\tilde n_B)^2/4u} \nn
    &= \frac{g^2 C_F}{4\pi^{2-\epsilon}}   
    \frac{ \left(b_\perp^2\right)^{\epsilon }\Gamma(1-\epsilon)}{2 \epsilon_{\rm IR} }
    \frac 12 
    \log \left(\frac{\left(r_a-1\right) \left(r_b-1\right)}{\left(r_a+1\right) \left(r_b+1\right)} \right)
    \frac{r_a r_b+1}{ \left(r_a+r_b\right)} \, .
    \label{eq:s1ab}
  \end{align}
  Thus, we obtain a finite result for diagram \ref{fig:Sa}. However, this result is only valid when $|r_a|,|r_b|>1$. This can be seen from expanding the term in the exponential of line 1 in Eq. (\ref{eq:s1ab}):
  \begin{align}
    (b_\perp+s\tilde n_A-t\tilde n_B)^2
    &= b_\perp^2 + s^2 (n_A^0)^2\left( r_a^2 -1 \right)
      + t^2 (n_B^0)^2\left( r_b^2 -1 \right)
      + 2st n_A^0 n_B^0 \left( r_a r_b +1 \right) 
      > 0 \, .
      \label{rabIneq}
  \end{align}
  Equation (\ref{rabIneq}) must be positive for the integral in Eq. (\ref{eq:s1ab}) to converge, which holds only when the $|r_a|, |r_b| > 1$. Because $s$ and $t$ can vary independently of each other, the terms on the RHS of Eq. \ref{rabIneq} must individually be greater than zero.

  The complete one-loop result in Euclidean space with complex directional vectors is:
  \begin{align}
    S(b_T,\epsilon,r_a,r_b)
    &= 1 + \frac{\alpha_s C_F}{2\pi}    
    \left(\frac{1}{\epsilon } + \log (\pi b_\perp^2 \mu_0^2 e^{\gamma_E})\right)
    \left\{
      2
      +
      \log
      \left(
        \frac{\left(r_a-1\right)\left(r_b-1\right)}
          {\left(r_a+1\right)\left(r_b+1\right)}
      \right)
      \frac{r_a r_b+1}{r_a + r_b}
    \right\} \, .
    \label{eq:ole}
  \end{align}

  The Collins scheme employs space-like Wilson lines, but in principle, time-like Wilson lines can also be used to regulate the rapidity divergence. For completeness, we present both the space-like and time-like cases here:
  \begin{align}
    \text{Time-like:}& \quad 
      n_A = \left( 1+e^{-2y_A}, \vec 0_\perp, 1-e^{-2y_A}  \right)
      , \quad 
      n_B = \left( 1+e^{2y_B}, \vec 0_\perp, -1+e^{2y_B}  \right) \\
    \text{Space-like:}& \quad 
      n_A = \left( 1-e^{-2y_A}, \vec 0_\perp, 1+e^{-2y_A}  \right)
      , \quad 
      n_B = \left( 1-e^{2y_B}, \vec 0_\perp, -1-e^{2y_B}  \right) \, . 
  \end{align}
  Since the complex directional vectors in Euclidean space are expressed in terms of the components of their Minkowski space counterparts, we define $r_a$ and $r_b$ using the components of the space-like or time-like directional vectors. For the time-like case, we obtain:
  \begin{align}
    r_a = \frac{1-e^{-2y_A}}{1+e^{-2y_A}}
    , \quad 
    r_b = \frac{1-e^{2y_B}}{1+e^{2y_B}} \, ,
  \end{align}
  where we can see that $-1<r_a,r_b<1$. We immediately find that this fails to satisfy the condition set by Eq. (\ref{rabIneq}). As for the space-like case, we find:
  \begin{align}
    r_a = \frac{1+e^{-2y_A}}{1-e^{-2y_A}}
    , \quad 
    r_b = \frac{1+e^{2y_B}}{1-e^{2y_B}} \, ,
    \label{eq:spaceR}
  \end{align}
  which satisfies Eq. (\ref{rabIneq}). 
  The third condition, set by the last term on the RHS of Eq. (\ref{rabIneq}), demands that $n_A^0 n_B^0 (r_a r_b +1) >0$, which indicates that the Wilson lines must be both future pointing or both past pointing, corresponding to $e^+e^-$ annihilation or DY type processes. 

  Substituting Eq. (\ref{eq:spaceR}) into Eq. (\ref{eq:ole}), we recover the Minkowski space result:
  \begin{align}
    S(b_\perp, \epsilon, y_A, y_B)
    &=
    1 + \frac{\alpha_s C_F}{2\pi}
    \left( \frac 1{\epsilon} 
      + \ln \left(\pi b_\perp^2 \mu_0^2e^{\gamma_E} \right) \right)
    \left\{
      2 -
      2 |y_A - y_B|
      \frac{1+e^{2(y_B-y_A)}}{1-e^{2(y_B-y_A)}}
    \right\}
    + \mathcal O(\alpha_s^2) \, .
  \end{align}
  Since $r_{a,b}$ directly map to the rapidities $y_{A,B}$, one can compute the soft function for several values of $r_{a,b}$ and fit the resulting plot to obtain the intrinsic soft function, $S_I$ in:
  \begin{align*}
    S\left( b_\perp, y_A, y_B, \mu\right) 
    \overset{\substack{y_A\to+\infty \\ y_B\to -\infty}}{=} S_I\left( b_\perp, \mu\right)  
      e^{2K\left( b_\perp, \mu\right) (y_A - y_B)} \, ,
  \end{align*}
  and the CS kernel:
  \begin{align}
    K(\mu, b_\perp)
    &=
    \lim\limits_{\substack{y_A\to+\infty \\ y_B\to -\infty}}
    \frac 12 \pdv{y_n} \log \left( \frac{S(b_\perp, y_n, y_B, \mu)}{S(b_\perp, y_A, y_n, \mu} \right)  \, .
  \end{align}

  \subsection{Finite Wilson lines}

  The lattice calculation is constrained by its finite space-time volume, meaning that Wilson lines must have a finite length $L$. To account for this, we also perform a one-loop calculation for Wilson lines of finite length.
  One consequence of finite-length Wilson lines is the appearance of a linear divergence in $L$. This divergence can be eliminated by constructing the ratio (\cite{collins2011b,jiliuliu2020}):
  \begin{align}
    S\left(b_\perp,a,r_a,r_b\right)
    &=
    \lim_{L\to\infty}
    \frac{S\left(b_\perp,a,r_a,r_b,L\right)}
      {\sqrt{S\left(b_\perp,a,r_a,-r_a,L\right)S\left(b_\perp,a,-r_b,r_b,L\right)}} \, ,
    \label{eq:ratio}
  \end{align}
  where $a$ is some generic regulator used for the UV divergence. 

  We have determined that Eq. (\ref{eq:ratio}) holds to one loop in perturbation theory, using a Polyakov regulator for the UV. Additionally, we find that power corrections of order $b_\perp^2/L^2$ will cancel in the ratio.
  By computing the ratio in Eq. (\ref{eq:ratio}) on the lattice for sufficiently large $L$, we expect to obtain a time-independent observable with power-suppressed corrections of order $b_\perp^4/L^4$, since $L$ in the lattice formulation is proportional to Euclidean time.

\section{Strategy of numerical implementation}

  It is well established (\cite{gervaisnevau1979,arefeva1980}) that the Wilson line can be expressed in terms of a one-dimensional auxiliary fermion field that "travels" along the path of the Wilson line: 
  \begin{align}
    P \exp 
      \left\{-ig\int_{s_i}^{s_f} \dd s n^\mu A_\mu(y(s)) \right\}
    &= 
    Z_{\psi}^{-1} \int \mathcal D \psi \mathcal D \bar \psi \,
      \psi \bar\psi
      \exp
      \left\{
        i \int_{s_i}^{s_f} \dd s \bar\psi i n\cdot\partial \psi 
        - g_0 \bar\psi n \cdot A \psi
      \right\} \, , 
  \end{align}
  whose propagator, $H_n(x-y)$, satisfies the Green function equation: 
  \begin{align}
    in\cdot D H_n(x-y) = i\delta^{(4)}(x-y) \, ,
    \label{eq:grM}
  \end{align}
  where $D=\partial+ig_0 A$ is the covariant derivative.
  The Euclidean space counterpart of {Eq. (\ref{eq:grM})} is defined with a directional vector with a purely imaginary time component, which can be written as $\tilde n = (in^0,\vec n)$ in terms of the Minkowski space vector components. After applying a Wick rotation, the Euclidean space Green function becomes: 
  \begin{align}
    i \tilde n \cdot D_E H_{\tilde n} (x_E-y_E) = \delta^{(4)}(x_E-y_E) \, .
    \label{eq:greenE}
  \end{align}
  As discussed in \cite{agliettieetal1992, Aglietti:1993hf}, meaningful solutions to Eq. (\ref{eq:greenE}) can only be obtained with a UV cutoff. Hence, using the lattice spacing as our regulator, we can construct a discretized version of this propagator.
  The central idea is to express the soft function in terms of lattice-regulated auxiliary field propagators, $H_{\tilde n}$, which can be shown to possess a well-defined continuum limit.

  Solving the auxiliary field propagator on the lattice follows the same procedure as in mHQET \cite{Mandula:1990fit, mandulaogilvie1992, agliettieetal1992, Aglietti:1993hf}, but now without the kinematic restrictions associated with heavy quarks. Consequently, we employ the recursive relation provided in Eq. (31) of \cite{mandulaogilvie1992} to construct the lattice auxiliary field propagators. Based on our perturbative analysis, we anticipate that computing Eq. (\ref{eq:ratio}) on the lattice will approach a time-independent result as the Euclidean time becomes large enough.
  
  \begin{figure}[t]
    \centering
    \begin{subfigure}{0.49\textwidth}
      \savebox{\tempfig}{\includegraphics[scale=\myscale,left]{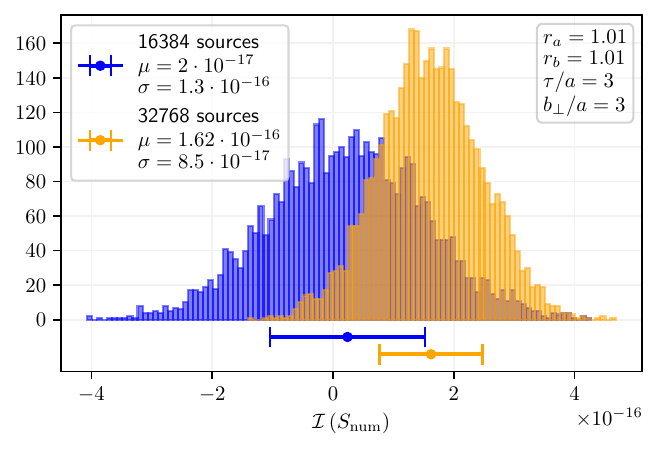}}
      \raisebox{\dimexpr\ht\tempfig-\height}{\includegraphics[scale=\myscale,right]{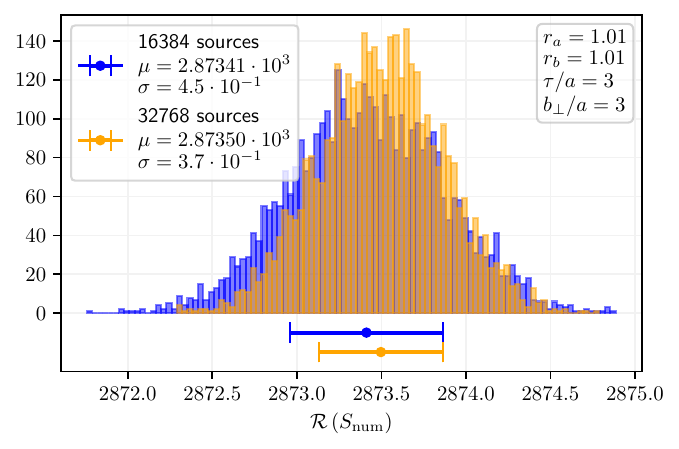}}
    \end{subfigure}
    \begin{subfigure}{0.49\textwidth}
      \includegraphics[scale=\myscale,left]{images/plots/hist_num_imag_1.01.pdf}
    \end{subfigure}
    \caption{Real and imaginary parts of the numerator loop, $S_{\rm num}$. The mean, $\mu$, is taken directly from the lattice measurement, and $\sigma$ is the standard deviation of the bootstrap distribution.}
    \label{fig:num}
  \end{figure}
  \begin{figure}[t]
    \centering
    \begin{subfigure}{0.49\textwidth}
      \includegraphics[scale=\myscale,right]{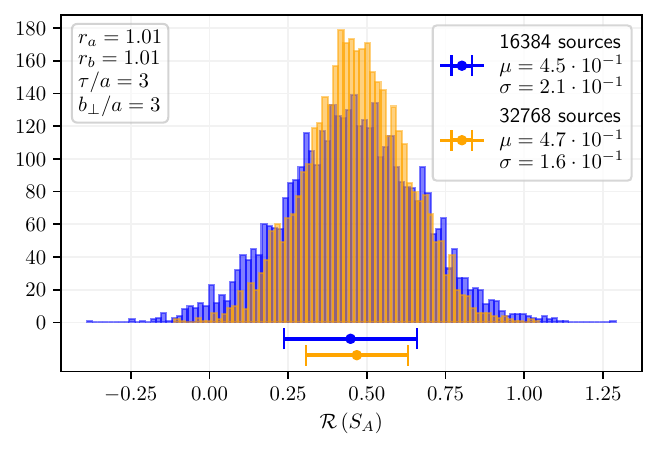}
    \end{subfigure}
    \begin{subfigure}{0.49\textwidth}
      \includegraphics[scale=\myscale,left]{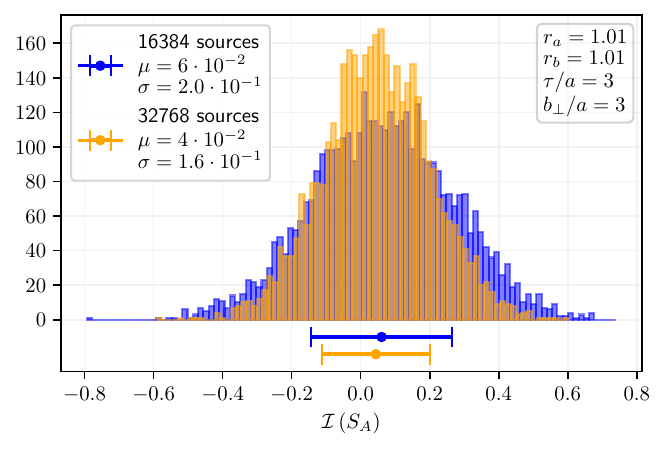}
    \end{subfigure}
    \caption{Real and imaginary parts of the denominator A loop, $S_{A}$. The mean, $\mu$, is taken directly from the lattice measurement, and $\sigma$ is the standard deviation of the bootstrap distribution.}
    \label{fig:denA}
  \end{figure}
  \begin{figure}[t]
    \centering
    \begin{subfigure}{0.49\textwidth}
      \includegraphics[scale=\myscale,right]{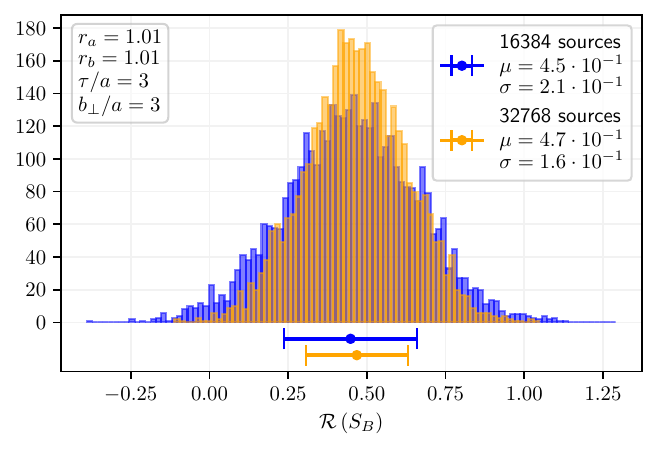}
    \end{subfigure}
    \begin{subfigure}{0.49\textwidth}
      \includegraphics[scale=\myscale,left]{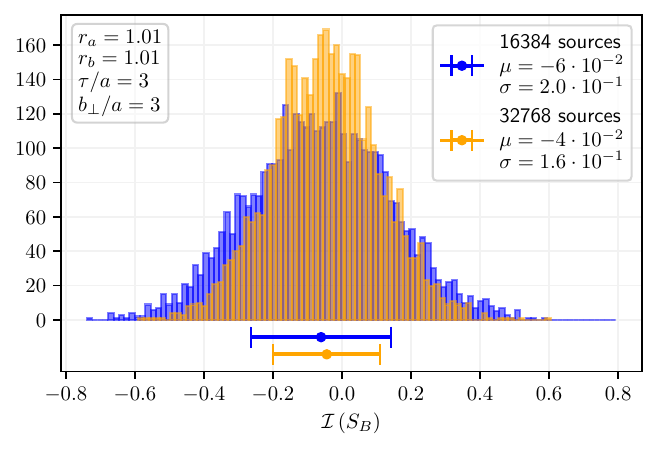}
    \end{subfigure}
    \caption{Real and imaginary parts of the denominator B loop, $S_B$. The mean, $\mu$, is taken directly from the lattice measurement, and $\sigma$ is the standard deviation of the bootstrap distribution.}
    \label{fig:denB}
  \end{figure}
  \begin{figure}[t]
    \centering
    \begin{subfigure}{0.49\textwidth}
      \centering
      \savebox{\tempfig}{\includegraphics[scale=\myscale,left]{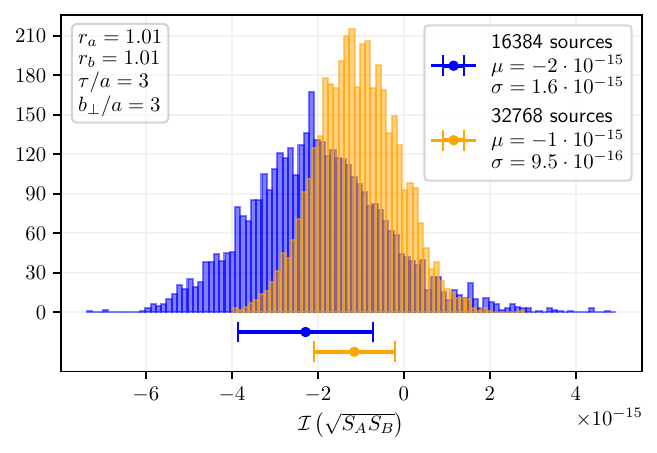}}
      \raisebox{\dimexpr\ht\tempfig-\height}{\includegraphics[scale=\myscale,right]{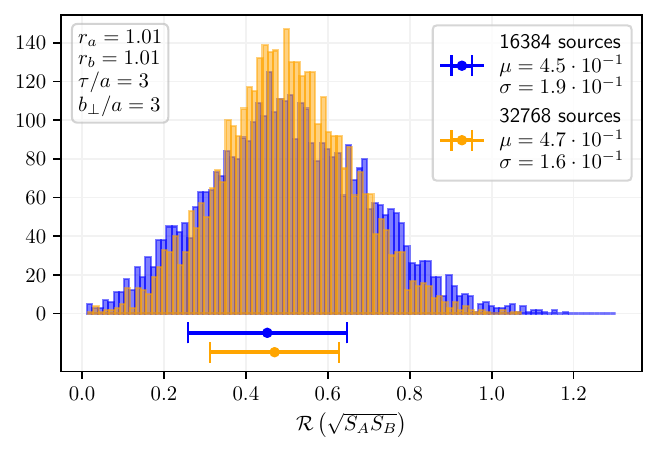}}
    \end{subfigure}
    \begin{subfigure}{0.49\textwidth}
      \centering
      \includegraphics[scale=\myscale,left]{images/plots/hist_den_imag_1.01.pdf}
    \end{subfigure}
    \caption{Real and imaginary parts of full denominator term, $\sqrt{S_A S_B}$. The mean, $\mu$, is taken directly from the lattice measurement, and $\sigma$ is the standard deviation of the bootstrap distribution.}
    \label{fig:den}
  \end{figure}

\section{Exploratory numerical results and discussion}

  For the lattice computation we use 400 dynamical configurations on a $32^3 \times 64$ lattice with $a=0.0907(13)~{\rm fm}$, light quark mass, $m_{\rm ud} = 2.527(47)~{\rm MeV}$, and strange quark mass, $m_{\rm s} = 72.72(78)~{\rm MeV}$, generated by the PACS-CS Collaboration \cite{PACS-CS:2008bkb}. 
  These are $N_f = 2 + 1$ flavor configurations using a non-perturbatively $\mathcal O(\alpha)$-improved Wilson quark action and Iwasaki gauge action. For our computation, we used two steps of THYP smearing \cite{Hasenfratz:2001hp}. 

  Our preliminary numerical exploration gives promising results, but highlights a clear need for increased statistics and a more involved analysis. 
  Based on a naive tree level computation, we expect the numerator and denominator factors to behave according to:
  \begin{align}
    S_{\rm num} \overset{\tau\to\infty}{\sim} e^{2 \tau (r_a+r_b)/a}/\tau^4 ,\nn
    S_{A} \overset{\tau\to\infty}{\sim} e^{4(\tau r_a - i z(\tau, r_a))/a}/\tau^4 ,\nn
    S_{B} \overset{\tau\to\infty}{\sim} e^{4(\tau r_b + i z(\tau, r_b))/a}/\tau^4 ,
    \label{eqn:treeb}
  \end{align}
  at large lattice time, $\tau$. The presence of an uncanceled oscillatory factor in the two denominator loops is a result of the cutoff effects particular to the auxiliary propagator in Euclidean space. While concerning, these contributions cancel upon taking the product of the two in $\sqrt{S_A S_B}$. Furthermore, upon constructing the desired ratio we obtain:
  \begin{align}
      S_{\rm ratio} = \frac{S_{\rm num}}{\sqrt{S_A S_B}} \overset{\tau\to\infty}{\sim} { \rm constant } \, ,
  \end{align}
  so we expect to see these unphysical contributions (cutoff effects) from the Euclidean space auxiliary field propagator cancel in the ratio.

  Given the large amount of noise present in gluonic observables, worsened in our case due to the oscillatory factors in  Eq. \ref{eqn:treeb}, we require a large number of source points per configuration in order to obtain precise results. At current statistics, we have 32,768 sources and we are able to demonstrate a clear increase in the statistical precision of our computation. Figures (\ref{fig:num}-\ref{fig:den}) demonstrate the results from a lattice computation at fixed lattice time $\tau/a = 3$, transverse separation $b_\perp/a = 3$, and $r_{a},r_{b} = 1.01$. 
  
  Looking at the bootstrap distribution in Fig. \ref{fig:num}, we see a comparison between the numerator loop with 32,768 sources and 16,384 sources, and a clear increase in precision when the number of sources is doubled. 
  The imaginary part is negligible in this case (note the scale of the x-axis in the imaginary plot in Fig. \ref{fig:num}), consistent with our naive expectation in Eq. \ref{eqn:treeb}.
  The denominator loops also show a clear increase in precision with increasing number of sources, but the relative error is much larger due to the presence of the oscillatory factor in Eq. \ref{eqn:treeb}.
  Combining the denominator loops into $\sqrt{S_A S_B}$ in Fig. \ref{fig:den}, we see that the imaginary part is negligible relative to the real part, also consistent with our expectations from Eq. \ref{eqn:treeb}. Also, comparing Figs. (\ref{fig:denA}) and (\ref{fig:denB}), the results are conjugate to each other.
  
  Taking the ratio in Eq. \ref{eq:ratio} with our current values for the numerator and denominator loop results in the the quotient: $2873.5/0.47$, much larger than one might expect. 
  A possible explanation for this is the need for the perturbative matching between the lattice and continuum regularization scheme.
  Nonetheless, in order to obtain a meaningful result on the configurations we are currently using, we require a far greater number of sources.

\section{Conclusions}
  
  The auxiliary field approach to extracting the soft function (what we compute in Eq. (\ref{eq:ratio})), intrinsic soft function, and CS kernel appears promising, due to the time-independence of the soft function and the correspondence between the complex directional vectors $\tilde n_A, \tilde n_B$ in Euclidean space and $y_A, y_B$ in Minkowski space in our perturbative analysis.
  One crucial aspect not addressed here is that the direct analytic continuation of our lattice calculations to Minkowski space could be complicated by the pole structure of the auxiliary field propagator. Nonetheless, this issue has been analyzed previously in the context of mHQET, as discussed in \cite{Aglietti:1993hf}.
  Finally, constructing the ratio in Eq. (\ref{eq:ratio}) effectively addresses the linear divergence in $L$ and eliminates power corrections at least up to $b_\perp^2/L^2$.
  Based on our current numerical results, we are able to demonstrate improvements in precision with increased statistics, while it remains to be seen whether we can obtain a clear signal for the Wilon loops in the denominator of Eq. (\ref{eq:ratio}). 
  In future works, we will explore the results of various smearing techniques used to enhance the signal at the same amount of statistics.
  \footnote{ 
  After our previous proceeding \cite{Francis:2023} was published, we were made aware of a lecture note~\cite{Liu:2022nnk} which implies that one can analytically continue the heavy-quark form factor~\cite{jiliuliu2020} in the $z$-axis to obtain Euclidean Wilson lines along the $(n^0,0,0,in^3)$ direction. This would be equivalent to our proposal through O(4) symmetry~\cite{Liu:2022nnk} if $n^3<n^0$. Nevertheless, such an analytical continuation was not explicitly pointed out in Refs.~\cite{jiliuliu2020,Liu:2022nnk}, nor was there mention of the failure of the standard timelike moving HQET in calculating the soft function, which was discovered in our work.
  } 

\acknowledgments

  We acknowledge Yizhuang Liu for helpful discussions.
  AF is supported by the National Science and Technology Council (NSTC) of Taiwan under grant 111-2112-M-A49-018-MY2. CJDL is supported by 112-2112-M-A49-021-MY3, and WM is supported by 112-2811-M-A49-517-MY2 and 112-2112-M-A49-021-MY3 from NSTC. YZ is supported by the U.S. Department of Energy, Office of Science, Office of Nuclear Physics through Contract No.~DE-AC02-06CH11357, and the 2023 Physical Sciences and Engineering (PSE) Early Investigator Named Award program at Argonne National Laboratory.

\end{document}